\documentclass[prd,nofootinbib,shownopacs,preprint]{revtex4}
\usepackage{graphicx}
\usepackage{epsfig}
\usepackage{amsmath}
\usepackage{amsfonts}
\usepackage{amssymb}
\usepackage{url}
\usepackage{hyperref}
\usepackage{subfigure}
\usepackage{hhline}
\usepackage{color}
\usepackage{bm}
\usepackage{cancel}
\usepackage{xspace}
\usepackage{blindtext}
\usepackage[italic]{hepnames}
\usepackage{amsmath,mathtools}
\usepackage{booktabs}

\newcommand{\be}{\begin{equation}}
\newcommand{\ee}{\end{equation}}
\newcommand{\bea}{\begin{eqnarray}}
\newcommand{\eea}{\end{eqnarray}}

\newcommand{\Bbar}{\,\overline{\!B}{}}
\newcommand{\Dbar}{\,\overline{\!D}{}}
\newcommand{\Kbar}{\,\overline{\!K}{}}
\def\B0bar{\Bbar{}^0}
\def\D0bar{\Dbar{}^0}
\def\K0bar{\Kbar{}^0}

\usepackage[toc,page]{appendix}
\usepackage{comment}
\begin{document}
\title{Investigating $ \Upsilon(ns) \to \tau^+ \tau^- $ decay in the Leptoquark scenario}
\author{Sambit Kumar Pusty}
\email{pustysambit@gmail.com}

\author{Dhiren Panda}
\email{pandadhiren530@gmail.com}

\author{Aishwarya Bhatta}
\email{aish.bhatta@gmail.com}

\affiliation{School of Physics,  University of Hyderabad, Hyderabad-500046,  India
}
\begin{abstract}
 Several measurements on $R_D, R_{D^*}$, and $R_{J/ \psi}$ by the BaBar, Belle, and LHCb experiments show significant deviations from their Standard Model (SM) predictions, which illustrate the fact that the concept of lepton flavor universality (LFU) is violated in semileptonic $B$ meson as well as leptonic $\Upsilon(ns)(n=1,2,3)$ decays. Recently BaBar experiment announced that at $1.8\sigma$ level, $R_\Upsilon(3s)= {\rm Br}(\Upsilon(3s) \to \tau \Bar{\tau})/{\rm Br}(\Upsilon(3s) \to \mu \Bar{\mu})$ shows an acceptance with the SM.  These fascinating findings point towards the possible implication of new physics in the $b \to c \tau \Bar{\nu}$ transitions,  which in turn, creates a new direction to look for new physics in $b \Bar{b} \to \tau \Bar{\tau} $ process. Thereby, the new physics contributions to the $b \to c \tau \Bar{\nu}$ process would inevitably alter the $b \Bar{b} \to \tau \Bar{\tau} $ transitions. Here we conduct a $\chi^{2}$ fit to constraining the new parameters by using the measured values of  $ R_D$, $R{_D*}$, $ R_{J/ \psi}$, $ R_{X_C}$, $F_L (D^*)$ and $P_{\tau} (D^*)$. In this context, we investigate the effect of constrained new physics couplings on the branching ratios and LFU parameters $R_{\Upsilon(ns)}$ through leptoquark models such as $S_3, \Tilde{S_1}, \Tilde{R_2}, U_1, U_3  $ and $ V_2$.
\end{abstract}

\maketitle
\section{Introduction}

The Standard Model (SM) of particle physics is the most successful theory that describes the basic interactions of the  elementary particles and their interactions. Numerous experiments have supported this conclusion. The so-called Lepton Flavour Universality (LFU) is one of the key characteristics of the SM that has been widely investigated over the past few years.\\
 Apart from phase-space differences or helicity suppression effects, the Standard Model (SM) estimates that the amplitudes of electroweak interactions will be identical across all three lepton generations, which are referred to as Lepton Flavour Universality. However, in the last few years, several $B$ meson  decays have revealed hints of deviations from LFU, intrigued the public's interest, and prompted investigations for lepton flavor-violating processes.
 
The $B$ meson phenomenology provides an excellent platform for critically scrutinizing  the Standard Model (SM) and exploring new physics (NP) effects  on  theoretical as well as experimental levels. Any discrepancy between theoretical and experimental estimations may reflect lepton universality violation and subsequently opens up a clear path to the new physics.

For the first time, the BABAR Collaboration  \cite{BABAR1,BABAR2} reported a discrepancy in the measurements of the semileptonic $B$ decays relative to the SM in the values of $R(D^{(*)}) $, defined as
\begin{equation}
  R(D^{(*)}) \equiv \frac{{\rm Br}( B \to D^{(*)} \tau\Bar{\nu}_\tau)}{{\rm Br}(  B \to D^{(*)} l \Bar{\nu_l})}\;.  
\end{equation}

The measurement of the above ratios in semileptonic $B$ meson decays has been performed by the experiments BABAR, Belle, and LHCb, which conclusively  point towards the possible signal  of lepton flavor universality violation (LFUV).

The Heavy Flavor Averaging Group (HFLAV) recently published the most recent global average values for 2022, that were analyzed using information recorded from the Belle, LHCb, and BABAR collaborations for $R(D^*) $ and $R(D)$ measurements  as
\begin{equation}
    \begin{split}
        R(D^*) =0.285 \pm 0.010 \pm 0.008\;, ~~~~           
        R(D) =0.358 \pm 0.025 \pm 0.012 \;,           
    \end{split}
\end{equation}
which show the deviations at  the level of $\sim 3 \sigma$, from the corresponding SM values:
\begin{equation}
    \begin{split}
        R(D^*) 
             = 0.254 \pm 0.005\;,~~~~~
        R(D) = 0.298 \pm 0.004\;.
    \end{split}
\end{equation}

Here, the most recent results are drawn from \cite{BABAR3,BABAR4,belle1,lhcb1,belle2,belle3,lhcb3,lhcb4,belle4} and HFLAV \cite{web1}.
All these measurements suggest LFU violation and are usually  regarded as LFU violating anomalies in  the charged current mediated semileptonic decays $b \to c \tau \bar \nu$. 

Considering the model-dependent framework, that takes the massive leptoquarks into account is one intriguing way to account for the anomalies. Leptoquarks have emerged as one of the fascinating candidates for explaining the observed anomalies. It is thought that these massive particles have the same number of generations as the leptons and quarks.
These massive bosons  can couple to both  quarks and  leptons simultaneously and can be found in different extensions of the Standard Model.

NP contributions to the $b \to c l \Bar{\nu_l}$ transition would  imply that NP contributions to the $b\bar{b}\to l^+ l^-$  transitions \cite{faroughy1, duque} are inevitable, given the SM content and the assumption that neutrinos are solely left-handed.
At the same time, the ratio of bottomonium meson $\Upsilon( b \Bar{b})$ leptonic decays can be used to test LFU\cite{iguro2}.
The fact that all processes take place at the same energy scale makes it favorable to use the correlation between the  operators involved in $ R(D^*)$ and those altering $\Upsilon(b \Bar{b})$ leptonic decays. Analogous to $ R(D^*)$, the LFU violating operator in these decays is defined as
\begin{equation}
 R(\Upsilon(ns)) \equiv \frac{{\rm Br}( \Upsilon (ns) \to \tau^+ \tau^-)}{{\rm Br}( \Upsilon(ns) \to l^+ l^-)} \;.   
\end{equation}
Here, $n= 1,2,3$, denotes various excited states (resonances) of the $\Upsilon$  mesons establishing a vibrant theoretical perspective.

The experiments conducted by BABAR and CLEO collaborations to inspect the LFU by measuring $R(\Upsilon)$'s \cite{babar5,cleo1} are shown in Table 1 below with the SM predictions \cite{aloni}.
\begin{table}[h]
\centering
{
\begin{tabular}{|c|c|c|c|}

\hline
 $\Upsilon(nS)$  & SM results        & Experimental values & Deviation \\ \hline 
$\Upsilon(1S)$ & $0.9924 \pm \mathcal{O}(10^{-5}$) &$1.005\pm0.013\pm 0.022$ & $0.5 \sigma$\\\hline 
$\Upsilon(2S)$ &$ 0.9940 \pm \mathcal{O}(10^{-5}$) &$1.04\pm0.04\pm0.05$ & $0.8 \sigma$\\\hline 
$\Upsilon(3S)$ & $0.9948 \pm \mathcal{O}(10^{-5}$) &$1.05\pm0.08\pm0.05 $& $0.6 \sigma$\\ \hline
\end{tabular}
\caption{SM predictions and the outcomes of the experiment for $ R({\Upsilon} )$}
}
\end{table}

The relevant masses of these resonant states with greater accuracy are listed below \cite{pat}.
\begin{center}
$m_{\Upsilon(1S)}$=9.46030 $\pm$ 0.00026 GeV,
\\$m_{\Upsilon(2S)}$=10.02326 $\pm$ 0.00031 GeV,
\\$m_{\Upsilon(3S)}$=10.3552 $\pm$ 0.0005 GeV,
\\$m_{\tau}$=1.77686 $\pm$ 0.00012 GeV.
\end{center}
This paper is organized in the following manner. We present the theoretical framework for the leptonic decays of $\Upsilon$ in section  II, where we Considered the most general effective Hamiltonian describing the decay process  $V \to l^+ l^-$ and obtained the expressions for the branching fraction as well as  the lepton non-universality parameter. To show the generalized new Physics scenario, in section III, we provide the details of leptoquarks we used  in $\Upsilon(ns) \to \tau^+ \tau^-$ processes. In section IV, we summarise our  results and discuss about the conclusion of our work. 
  
\section{Theoretical Framework}
The most general effective Lagrangian describing the $\Upsilon \to l^+l^-$ decay processes is given as
\begin{equation}
    \begin{split}
        \mathcal L_{lq} & = C_{V_{RR}}\bar{e}_R \gamma^\mu e_R \bar{q}_R \gamma_\mu q_R +C_{V_{RL}}\bar{e}_R \gamma^\mu e_R \bar{q}_L \gamma_\mu q_L +C_{V_{LR}}\bar{e}_L \gamma^\mu e_L \bar{q}_R \gamma_\mu q_R  \\ &
        +C_{V_{LL}}\bar{e}_L \gamma^\mu e_L \bar{q}_L \gamma_\mu q_L +[C_T\bar{e}_L \sigma^{\mu\nu} e_R \bar{q} \sigma_{\mu\nu} q+C_{S_L}\bar{e}_R  e_L \bar{q}_R q_L+C_{S_R}\bar{e}_R  e_L \bar{q}_L q_R \\& +h.c.],
    \end{split}
\end{equation}
where $ C_{V_{LL}},~C_{V_{LR}},~C_{V_{RL}},~C_{V_{RR}}, ~C_{T}, ~C_{S_L},~C_{S_R} $ are the NP couplings.

 It is possible to express the generalized decay width and the LFU parameter as
\begin{equation}
\begin{split}
      \Gamma &=\frac{f_v^2}{4\pi M_v} \sqrt{1-4x_l^2}[| A_v |^2(1+2x_l^2)+2~Re [A_v~C_v]x_l(1-4x_l^2)]+| B_v |^2(1-4x_l^2)\\
      & +\frac{| C_v|^2}{2}(1-4x_l^2)  +\frac{|D_v|^2}{2}(1-4x_l^2)\;,
\end{split}
\end{equation}
and
\begin{equation}
    \begin{split}
    R_{\tau/l}^V &=\frac{\sqrt{1-4x_l^2}}{| A_v^{SM} |^2} [| A_v |^2(1+2x_l^2)+2~Re [A_v~C_v]x_l(1-4x_l^2)]+|B_v|^2(1-4x_l^2) \\&
    +\frac{| C_v |^2}{2}(1-4x_l^2) 
    +\frac{| D_v|^2}{2}(1-4x_l^2) \;,
    \end{split}
\end{equation}
where,
\begin{equation}
\begin{split}
A_v &=-4 \pi \alpha Q_q+\frac{m_v^2}{4}[(C_{V_{LL}}+C_{V_{RR}}+C_{V_{LR}}+C_{V_{RL}})+16 x_l \frac{f_V^T}{f_V} Re[C_T]],\\
B_v &=\frac{m_v^2}{4}[(C_{V_{LL}}+C_{V_{RR}}-C_{V_{LR}}-C_{V_{RL}}],\\
C_v &= 2 m_V^2 \frac{f_V^T}{f_V} Re[C_T],\\
D_v &= 2 m_V^2 \frac{f_V^T}{f_V} Im[C_T],\\
\end{split}
\end{equation}
with $ x_l=\frac{m_l}{m_V}$ and $f_V,~~ f_V^T$ are the form factors. Within the Standard Model, $A_v =-4 \pi\alpha Q_q$ and other terms such as $ B_v,~ C_v,~D_v$=0. The form factor in the Heavy quark limit is taken to be $f_V=f_V^T$, and the value of the form factor is taken from Ref \cite{duque1}.

\section{A General  Leptoquark  Scenarios}

Leptoquarks (LQs) are hypothetical particles that can turn quarks into leptons and viceversa of either scalar or vector nature \cite{Shanker:1981mj,Shanker:1982nd,Buchmuller:1986iq,Buchmuller:1986zs,Hewett:1987yg,Leurer:1993em,Leurer:1993qx,dor, Crivellin1, Crivellin2, Crivellin3}. Leptoquark discovery would be a tentative signal of matter unification of weak and electromagnetic interactions. The SM effective Hamiltonian can change in the presence of leptoquarks, causing a notable deviation from the SM values. In local quantum field theories they can be of either scalar (spin zero) or vector (spin one) nature. The SM gauge group $SU(3)_C \times SU(2)_L \times U(1)_Y$, where $Y$ is hypercharge, is invariant and contains scalar and vector-relevant leptoquark states that may contribute to our transitions. For instance, $X  \equiv (3,1,2/3)$ transforms as triplet (singlet) under $SU(3)(SU(2))$, and with hypercharge= 2/3. LQs have renormalizable complex couplings to two SM fermions and can thus be probed in low-energy experiments. They generate contributions to electric and magnetic dipole moments of quarks and leptons as well as rare radiative decays of mesons and neutral anti-meson oscillations. We focus on the low-energy effects of a single LQ exchange and derive limits on the corresponding LQ couplings, assuming that they are the only significant contribution to the relevant observables beyond the SM Yukawas. The Minimal Flavor Violation hypothesis has been applied to LQ scenarios in Refs. \cite{csaki,david1,nikol,arca}. The weak multiplet mass eigenstates of LQs that have non-trivial $SU (2)$ representations can show mass splittings. Changing mass scales might theoretically decrease their contributions to low-energy observables. Large inter-multiplet mass splittings are, however, severely constrained in practice by electroweak precision observable.
\begin{table}[tbp]
\centering
\begin{tabular}{|c|cccc|}
\hline
$(SU(3),SU(2),U(1))$ &  Spin & Symbol & Type & $F$ \\
\hline
$(\mathbf{3},\mathbf{1},2/3)$ & 1 & $U_1$ & $LL$\,$(V^L_0)$, $RR$\,$(V^R_0)$, $\overline{RR}$\,$(V^{\overline{R}}_0)$ & $0$ \\
$(\mathbf{3},\mathbf{1},5/3)$ & 1 & $\tilde{U}_1$ & $RR$\,$(\tilde{V}^R_0)$ & $0$ \\
$(\mathbf{3},\mathbf{1},-1/3)$ & 1 & $\bar{U}_1$ & $\overline{RR}$\,$(\bar{V}^{\overline{R}}_0)$ & $0$ \\
$(\mathbf{3},\mathbf{3},2/3)$ & 1 & $U_3$ & $LL$\,$(V^L_1)$ & $0$ \\
$(\overline{\mathbf{3}},\mathbf{2},5/6)$ & 1 & $V_2$ & $RL$\,$(V^L_{1/2})$, $LR$\,$(V^R_{1/2})$ & $-2$ \\
$(\overline{\mathbf{3}},\mathbf{2},-1/6)$ & 1 & $\tilde{V}_2$ & $RL$\,$(\tilde{V}^L_{1/2})$, $\overline{LR}$\,$(\tilde{V}^{\overline{R}}_{1/2})$ & $-2$ \\
\hline 
$(\overline{\mathbf{3}},\mathbf{1},4/3)$ & 0 & $\tilde{S}_1$ & $RR$\,$(\tilde{S}^R_{0})$ & $-2$ \\
$(\overline{\mathbf{3}},\mathbf{1},1/3)$ & 0 & $S_1$ & $LL$\,$(S^L_0)$, $RR$\,$(S^R_0)$, $\overline{RR}$\,$(S^{\overline{R}}_0)$ & $-2$ \\
$(\overline{\mathbf{3}},\mathbf{1},-2/3)$ & 0 & $\bar{S}_1$ & $\overline{RR}$\,$(\bar{S}^{\overline{R}}_0)$ & $-2$ \\
$(\overline{\mathbf{3}},\mathbf{3},1/3)$ & 0 & $S_3$ & $LL$\,$(S^L_1)$ & $-2$ \\
$(\mathbf{3},\mathbf{2},7/6)$ & 0 & $R_2$ & $RL$\,$(S^L_{1/2})$, $LR$\,$(S^R_{1/2})$ & $0$ \\
$(\mathbf{3},\mathbf{2},1/6)$ & 0 & $\tilde{R}_2$ & $RL$\,$(\tilde{S}^L_{1/2})$, $\overline{LR}$\,$(\tilde{S}^{\overline{L}}_{1/2})$ & $0$ \\

\hline

\hline 
\end{tabular}
\caption{\label{tab:LQs} List of vector and scalar LQs.}
\end{table} 
In comparison to the mechanism of rare SM decay, the LQ contributions to rare meson decay differ qualitatively. By rare decays, we mean those that only occur in loop diagrams, are suppressed by the unitarity of the CKM matrix (GIM mechanism), or even don't occur at all (lepton flavor violation) in the SM. These decays have neutral quark and lepton currents $(\Bar{q}'q)(\Bar{l}'l)$ or $(\Bar{q}'q)(\Bar{\nu}' \nu)$. The flavor-changing LQ couplings are severely constrained by the conflicting LQ amplitudes, which can be induced at the tree level.

The semileptonic charged current processes which are based on the direct tree-level probes the LQ effects, whereas leptonic charged processes proceed through higher perturbative order which results in diminished sensitivity to LQ couplings. The charged current processes can be instigated by conserving the baryon number of LQ states $\tilde R_2$, $U_3$, and $U_1$, containing a $2/3$ charged state. On the other
side, the baryon number violating states $S_3$, $\tilde S_1$ and $V_2$ entail charged currents via exchange of $Q=1/3$. In the presence of one of the LQs listed above, the generalization of the effective weak Lagrangian is given as:
\begin{equation}
 \label{eq:ccLag}
    \begin{split}
      \mathcal{L}_{\Bar{u}^i d^j \Bar{l} \nu_k} &=-\frac{4 G_F}{\sqrt2}[(V_{ij}U_{lk}+g^L_{ij;lk})(\Bar{u}^i_L \gamma^\mu d^j_L)(\Bar{l}_L \gamma_\mu \nu^k_L)\\&
      +g^R_{ij;lk}(\Bar{u}^i_R \gamma^\mu d^j_R) (\Bar{l}_R \gamma_\mu \nu^k_R) \\&
      +g^{LL}_{ij;lk}(\Bar{u}^i_L d^j_R) (\Bar{l}_L  \nu^k_R)+h^{LL}_{ij;lk}(\Bar{u}^i_L \sigma^{\mu\nu} d^j_R) (\Bar{l}_L \sigma_{\mu\nu} \nu^k_R)\\&
       +g^{RR}_{ij;lk}(\Bar{u}^i_R d^j_L) (\Bar{l}_R  \nu^k_L)+h^{RR}_{ij;lk}(\Bar{u}^i_R \sigma^{\mu\nu} d^j_L) (\Bar{l}_R \sigma_{\mu\nu} \nu^k_L)\\&
      +g^{LR}_{ij;lk}(\Bar{u}^i_L d^j_R) (\Bar{l}_R  \nu^k_L)+g^{RL}_{ij;lk}(\Bar{u}^i_R  d^j_L) (\Bar{l}_L  \nu^k_R)]+h.c..
    \end{split}
\end{equation}

In Eq.~\eqref{eq:ccLag} $G_F$ is the Fermi coupling constant and the indices $i,j,k $ and $ \ell$ refer to the fermion mass eigenstates. The term proportional to $V_{ij} U_{\ell k}$ survives In the SM limit. Modifications of the left-handed currents, parameterized by $g^{L}$, are expected in the presence of LQs that transform either as singlets or triplets under $SU(2)$. On the other hand, the right-handed currents proportional to $g^{R}$ are distinct signature of weak singlet LQ states. Further operators involving chirality flipping currents are possible in the presence of LQ states and are parameterized by couplings
$g^{XY}_{ij;\ell k}$, where $X$ and $Y$ refer to chiralities of the
up-type quark and charged lepton, respectively.\\
The effective Lagrangian for the dimension six operators coming from the tree level LQ coupling for $\bar{q}q\bar{l}l'$ process:
\begin{equation}
    \begin{split}
        \mathcal L_{\bar{q}^iq^j\bar{l}l'} & =-\frac{4G_F}{\sqrt{2}}[c^{LL}_{ij;ll'}(\bar{q_L}^i\gamma^{\mu}q_L^j)(\bar{l_L}\gamma_{\mu}l'_L)+c^{RR}_{ij;ll'}(\bar{q_R}^i\gamma^{\mu}q_R^j)(\bar{l_R}\gamma_{\mu}l'_R) \\&
        +c^{LR}_{ij;ll'}(\bar{q_L}^i\gamma^{\mu}q_L^j)(\bar{l_R}\gamma_{\mu}l'_R)+c^{RL}_{ij;ll'}(\bar{q_R}^i\gamma^{\mu}q_R^j)(\bar{l_L}\gamma_{\mu}l'_L) \\&        
        +g^{LL}_{ij;ll'}(\bar{q_L}^i q_R^j)(\bar{l_L} l'_R)+h^{LL}_{ij;ll'}(\bar{q_L}^i \sigma^{\mu\nu} q_R^j)(\bar{l_L} \sigma_{\mu\nu} l'_R)  \\&
        +g^{RR}_{ij;ll'}(\bar{q_R}^i q_L^j)(\bar{l_R} l'_L) +h^{RR}_{ij;ll'}(\bar{q_R}^i \sigma^{\mu\nu} q_L^j)(\bar{l_R} \sigma_{\mu\nu} l'_L) \\&
        +g^{LR}_{ij;ll'}(\bar{q_L}^i q_R^j)(\bar{l_R} l'_L)+g^{RL}_{ij;ll'}(\bar{q_R}^iq_L^j)(\bar{l_L} l'_R)]+h.c..
    \end{split}
\end{equation}
With the exception of coefficients $C^{LR }$and $C^{RL}$, whose counterparts are absent from the charged-current effective Lagrangian, the aforementioned operator basis is analogous to the charged-current semileptonic basis
 We propose the following mapping from the above-mentioned ``chiral basis" of operators in equation (9) to the basis typically utilized in rare $B$ meson decays \cite{buchalla} in order to connect to the substantial body of phenomenological research on rare leptonic and semileptonic meson decays.
 \begin{equation}
     \begin{split}
         \mathcal L_{\bar{q}^iq^j\bar{l}l'}  &=-\frac{4G_F}{\sqrt{2}}\lambda_q[C_7^{ij}\mathcal{O}_7^{ij}+C_{7'}^{ij}\mathcal{O}_{7'}^{ij}+\sum_{X=9,10,S,P}(C_X^{ij;ll'}\mathcal{O}_X^{ij;ll'}+C_{X'}^{ij;ll'}\mathcal{O}_{X'}^{ij;ll'})  \\ & +C_T^{ij;ll'}\mathcal{O}_T^{ij;ll'}+C_{T5}^{ij;ll'}\mathcal{O}_{T5}^{ij;ll'}]+h.c..
     \end{split}
 \end{equation}
Each of the mentioned operators is sensitive to LQ effects
\begin{equation}
    \begin{split}
        \mathcal{O}_7^{ij}&=\frac{em_{q^j}}{(4\pi)^2}(\Bar{q}^i \sigma_{\mu\nu}P_R q^j)F^{\mu\nu}, ~~~~~\mathcal{O}_S^{ll'}=\frac{e^2}{(4\pi)^2}(\Bar{q}^i P_R q^j)(\Bar{l}l'),\\
        \mathcal{O}_9^{ij;ll'}&=\frac{e^2}{(4\pi)^2}(\Bar{q}^i \gamma^\mu P_L q^j)(\Bar{l}\gamma_\mu l'), ~~~~~\mathcal{O}_P^{ij;ll'}=\frac{e^2}{(4\pi)^2}(\Bar{q}^i P_R q^j)(\Bar{l} \gamma_5 l'),\\
         \mathcal{O}_{10}^{ij;ll'}&=\frac{e^2}{(4\pi)^2}(\Bar{q}^i \gamma^\mu P_L q^j)(\Bar{l}\gamma_\mu \gamma_5 l'), ~~~~~ \mathcal{O}_T^{ij;ll'}=\frac{e^2}{(4\pi)^2}(\Bar{q}^i \sigma_{\mu\nu} q^j)(\Bar{l} \sigma^{\mu\nu} l'),\\
         \mathcal{O}_{T5}^{ij;ll'}&=\frac{e^2}{(4\pi)^2}(\Bar{q}^i \sigma_{\mu\nu} q^j)(\Bar{l} \sigma^{\mu\nu} \gamma_5 l')
    \end{split}
\end{equation}
By switching the roles of $P_L$ and $P_R$, the set of operators with primes is related to the unprimed $(X')$ set. where $P_L(P_R)=\frac{1}{2}(1\mp \gamma_5)$. The standard operator basis has an overall CKM factor q that reads $ \lambda_q = V_{qj} V_{qi} $ in the case of $ d_j \rightarrow d_i  l^- l'^+$ processes and $ \lambda_q = V_{jq} V_{iq} $ for $ u_j \rightarrow u_i  l^- l'^+$ processes. This basis is appropriate for the SM loop contributions. The conventions are based on the effective Hamiltonian structure in rare $B$ decays \cite{buchalla}, where the top quark's SM contribution is the dominating one, and  $q = t$ is used. One often sets q = b for the short-distance contributions to rare charm decays. Contrary to the above convention, the basis commonly used in rare kaon decays \cite{pich,mes} is different. The external quark mass ($m_{q^j}$ in the transition $q^j \to q^i $ ), which is assumed to be the mass of the decaying quark, is included in the definition of $ \mathcal{O}_7^{ij}$. The guidelines for mapping
\begin{equation}
    \begin{split}
        C_{9,10} &=\frac{2\pi}{\alpha_{em}\lambda_q}(c^{LR}\pm c^{LL}), ~~~~ C_{9',10'}=\frac{2\pi}{\alpha_{em}\lambda_q}(c^{RR}\pm c^{RL}),\\
        C_{S,P} &=\frac{2\pi}{\alpha_{em}\lambda_q}(g^{LL}\pm g^{LR}), ~~~~ C_{S',P'}=\frac{2\pi}{\alpha_{em}\lambda_q}(g^{RL}\pm g^{RR}),\\
        C_{T,T5} &=\frac{2\pi}{\alpha_{em}\lambda_q}(h^{LL}\pm h^{RR}),
    \end{split}
\end{equation}
with flavor indices $ij; ll'$ implied.
\\
Comparing equations (5), (9) and from \cite{dor}, the new physics coupling terms:
\begin{equation}
    \begin{split}
        C_{V_{LL}} &=-\frac{4G_F}{\sqrt{2}}C^{LL}, ~~~~
         C_{V_{RR}} =-\frac{4G_F}{\sqrt{2}}C^{RR}, \\ 
          C_{V_{LR}} &=-\frac{4G_F}{\sqrt{2}}C^{RL},~~~~  
           C_{V_{RL}} =-\frac{4G_F}{\sqrt{2}}C^{LR} .
    \end{split}
\end{equation}
The values of these new couplings can be obtained by performing a $\chi^2$ analysis. For that
we consider the NP contribution of only one coupling at a time, setting all others to zero to perform the chi-square  fitting of the individual coupling. $\chi^2$ can be defined as
\begin{equation}
    \chi^2 =\sum_i \left(\frac{\mathcal{O}_{i}^{th}-\mathcal{O}_{i}^{exp}}{ \Delta \mathcal{O}_{i}}\right)^2
\end{equation}
where $\mathcal{O}_{i}$ represents the theoretical prediction of the observable and $\mathcal{O}_{i}^{exp}$ symbolize the measured central values of the observables and $(\Delta \mathcal{O}_{i})^2$=$(\Delta \mathcal{O}_{i}^{th})^2$ +$(\Delta \mathcal{O}_{i}^{exp})^2$ contain the $1\sigma$ error from the theory and experiment. We constrain the new coefficient related to the concern process from the $\chi^{2}$ fit of $ R{_D*}$,~$ R_D$,~$ R_{J/ \psi}$,~$ R_{X_C}$,~$F_L (D^*)$,~$P_{\tau} (D^*)$. The updated value for observables is taken from the \cite{duque}.
We estimated the relevant Wilson coefficient values as 
\begin{equation}
    \begin{split}
      C_{V_{LL}} &=0.0575,~~~ C_{V_{RL}}=-0.0581,
      \\
    C_{V_{LR}} &=-0.5809,~~~  C_{V_{RR}}=-0.3458.  
    \end{split}
\end{equation}
Now, using these obtained values, we proceed to find out the values of the decay width of $\Upsilon(nS) \to \tau^+\tau^-$ process and the LFU parameter in various LQ scenarios.
\subsection{\textbf{{ Leptoquark $S_3 (\bar{3},3,1/3)$}}}
 $S_3$ scalar's interactions with matter are described by the following two operators
\begin{equation}
      \mathcal{L}\supset +y^{LL}_{3ij}\Bar{Q}^{Ci,a}_L  \epsilon^{ab} (\tau^k S^k_3)^bc L_L^{j,c}
      +z^{LL}_{3ij}\Bar{Q}^{Ci,a}_L  \epsilon^{ab} ((\tau^k S^k_3)^\dagger)^{bc} Q_L^{j,c}+h.c.,
\end{equation}
where the matrices $y(z)$ explain the ability of LQ interaction with quark-lepton(quark-quark) pairs. $y^{LL}_{3}$, $z^{LL}_{3}$  are components of a complex $3\times 3 $ Yukawa coupling matrix with $z^{LL}_{3}$ .  $i,j=1,2,3 ~~and~~ a,b=1,2$ are flavor indices. $\tau^{k(1,2,3)}$ Pauli matrices, $\epsilon^{ab}=(i\tau^2)^{ab}$, and $S^k_3$ are elements that make up $S_3$ in $SU(2)$ space. 

The relevant NC interaction is given by $C_{V_{LL}}$ as
\begin{equation}
 \mathcal L_{NC}  =  C_{V_{LL}}\bar{e}_L \gamma^\mu e_L \bar{q}_L \gamma_\mu q_L.
\end{equation}
The above Lagrangian induces the $\Upsilon \to \tau \tau$ process. Thus, with Eqns (6), (7) and (16), we obtainthe values of branching fraction and the LFU observables which are listed in Table \ref{lqs}.  
\subsection{\textbf{{ Leptoquark $\Tilde{S_1} (\bar{3},1,4/3)$}}}
For $\Tilde{S_1}$, the coupling  to matter expressed as

\begin{equation}
    \mathcal{L}\supset +\Tilde{y}^{RR}_{1ij}\Bar{d}^{Ci}_R  \Tilde{S_1} e_R^j+\Tilde{z}^{RR}_{1ij}\Bar{u}^{Ci}_R  \Tilde{S_1^*} u_R^j+h.c..
\end{equation}

The only surviving term is the term containing coefficient  $C_{V_{RR}}$. We observe that in any basis $\tilde{z}^{RR}_{1}$ is an antisymmetric matrix. The corresponding effective Lagrangian for NC interaction has the form
\begin{equation}
 \mathcal L_{NC}  =  C_{V_{RR}}\bar{e}_R \gamma^\mu e_R \bar{q}_R \gamma_\mu q_R . 
\end{equation}
The obtained results from this Lagrangian are listed in Table \ref{lqs}.
\subsection{\textbf{{ Leptoquark $\Tilde{R}_2 (3,2,1/6)$}}}
Only two renormalizable terms, that describe interactions of $\tilde{R}_2$ with matter, are available for this LQ.
\begin{equation}
    \mathcal{L}\supset -\Tilde{y}^{RL}_{2ij}\Bar{d}^i_R  \Tilde{R_2^a} \epsilon^{ab} L_L^{j,b}+\Tilde{y}^{\overline{LR}}_{2ij} \Bar{Q}^{i,a}_L  \Tilde{R_2^a} \nu_R^j+h.c..
\end{equation}
  $\tilde{y}^{RL}_{2\,ij}$ and $\tilde{y}^{\overline{LR}}_{2\,ij}$ are then elements of arbitrary complex $3 \times 3$ Yukawa coupling matrices. The only surviving terms are $C_{V_{LR}}, C_{V_{RL}}$. The corresponding effective Lagrangian has the form
\begin{equation}
 \mathcal L_{lq}  = C_{V_{LR}}\bar{e}_L \gamma^\mu e_L \bar{q}_R \gamma_\mu q_R .
\end{equation}
From this Lagrangian, the obtained values of the observables are listed in Table \ref{lqs}.

\subsection{\textbf{{ Leptoquark $U_1 (3,1,2/3)$}}}
The non-chiral vector LQ is $U_1$. It is coupled to SM fermions through the following mechanisms.
\begin{equation}
    \mathcal{L}\supset +x^{LL}_{1ij}\Bar{Q}^{i,a}_L \gamma^\mu U_{1,\mu}L_L^{j,a}+x^{RR}_{1ij}\Bar{d}^{i}_R \gamma^\mu U_{1,\mu}e^j_R+x^{\overline{RR}}_{1ij}\Bar{u}^{i}_R \gamma^\mu U_{1,\mu} \nu^j_R+h.c..
\end{equation}

$x_{1ij}^{LL}$ and $x_{1ij}^{RR}$ are the gauge couplings.
\\ The relevant NC interaction containing 
 relevant Willson coefficient term($C_{V_{LL}},~C_{V_{RR}}$) in mass eigen basis
\begin{equation}
 \mathcal L_{NC}  = C_{V_{RR}}\bar{e}_R \gamma^\mu e_R \bar{q}_R \gamma_\mu q_R + C_{V_{LL}}\bar{e}_L \gamma^\mu e_L \bar{q}_L \gamma_\mu q_L  .
\end{equation}
The results obtained from the above Lagrangian are listed in Table \ref{lqs}.

\subsection{\textbf{{ Leptoquark $U_3 (3,3,2/3)$}}}
$U_3$ has couplings of the LL type and is a true vector LQ. The essential operator presents.
\begin{equation}
\label{eq:main_U_3}
     \mathcal{L}\supset +x^{LL}_{3ij}\Bar{Q}^{i,a}_L \gamma^\mu (\tau^k U^k_{3,\mu})^{ab} L_L^{j,b}+h.c.,
\end{equation}
where $U^k_3$  are $U_3$'s constituents in $SU(2)$ space, and In the Lorentz space, $\mu$ is an index; $\mu(0,1,2,3)$. Usually vector LQs are  associated with carriers of new interactions. This implies that $x^{LL}_{3\,ij}$ in Eq.~\eqref{eq:main_U_3} can be identified with gauge couplings that measure the strength of relevant interactions. These are flavor blind. If that is assumed one could accordingly write $x^{LL}_{3\,ij}=x \delta_{ij}$, where $x$ would represents the gauge coupling in question.
\\Following is the relevant NC interaction term containing the Wilson coefficient  $C_{V_{LL}}$ 
\begin{equation}
 \mathcal L_{NC}  =  C_{V_{LL}}\bar{e}_L \gamma^\mu e_L \bar{q}_L \gamma_\mu q_L . 
\end{equation}
We thus, obtain the values of the branching fraction and LFU observables, which are listed in Table \ref{lqs}.

\subsection{\textbf{{ Leptoquark $V_2 (\Bar{3},2,5/6)$}}}

$V_2$ is a non-chiral vector LQ and is coupled to the SM fermions in the following ways
\begin{equation}
    \mathcal{L}\supset +x^{RL}_{2ij}\Bar{d}^{Ci}_R \gamma^\mu V^a_{2,\mu} \epsilon^{ab} L_L^{j,b}+x^{LR}_{2ij}\Bar{Q}^{Ci,a}_L \gamma^\mu \epsilon^{ab} V^b_{2,\mu}  e_R^j+w^{LR}_{2ij}\Bar{Q}^{Ci,a}_L \gamma^\mu \epsilon^{ab} V^{a*}_{2,\mu}  u_R^j
    +h.c..
\end{equation}

The only surviving terms are $C_{V_{LR}}, C_{V_{RL}}$. The corresponding effective lagrangian for NC interaction has the form
\begin{equation}
 \mathcal L_{NC}  =  C_{V_{RL}}\bar{e}_R \gamma^\mu e_R \bar{q}_L \gamma_\mu q_L +C_{V_{LR}}\bar{e}_L \gamma^\mu e_L \bar{q}_R \gamma_\mu q_R .
\end{equation}
The values obtained for various observables are listed Table \ref{lqs}.

\section{Results and Conclusion }
\begin{table}[h]
\begin{tabular}{||c|c|c|c|c|c|c||}
 	\hline
 \toprule
\multicolumn{1}{c}{} & \multicolumn{3}{c}{\textbf{$ \Gamma(\Upsilon(nS) \to \tau^+ \tau^-) $}} & \multicolumn{3}{c}{$ R_{\tau/l}^{\Upsilon(nS)} $} \\
\cmidrule(rl){2-4} \cmidrule(rl){5-7}
\hline
\textbf{LQs} & {1S} & {2S} & {3S} & {1S} & {2S} & {3S} \\ \midrule
\hline
$ S_3 $ &   $ 3.3976\times 10^{-6} $ & $ 1.6206\times 10^{-6} $ & $ 1.1759\times 10^{-6} $ & 0.9949 & 0.9969 & 0.9979\\
\hline
$\tilde{S_1}$ &  $ 3.4452\times 10^{-6} $ & $ 1.6460\times 10^{-6} $ & $ 1.1956\times 10^{-6} $ & 1.0088 & 1.0126 & 1.0146\\
\hline
$\tilde{R_2}$ & $ 3.3787 \times 10^{-6} $ & $ 1.6104\times 10^{-6} $ & $ 1.1680\times 10^{-6} $ & 0.9999 & 0.9999 & 0.9999\\
\hline
$ U_1 $ &  $ 3.4358\times 10^{-6} $ & $ 1.6410\times 10^{-6} $ & $ 1.1908\times 10^{-6} $ & 1.0060 & 1.0095 & 1.0105\\
\hline
$ U_3 $ &  $ 3.3788\times 10^{-6} $ & $ 1.6105\times 10^{-6} $ & $ 1.1681\times 10^{-6} $ & 0.9894 & 0.9907 & 0.9913\\

\hline
$ V_2 $ &  $ 3.4072\times 10^{-6} $ & $ 1.6257\times 10^{-6} $ & $ 1.1799\times 10^{-6} $ & 0.9977 & 1.0001 & 1.0013\\
\hline
 \bottomrule
\end{tabular}
\caption {\label{lqs} Observables in the presence of leptoquarks}
\end{table}
\paragraph{}
This section is dedicated to studying the LFV observable of the decay channel of the upsilon meson ($\Upsilon(ns)$), $(n=1,2,3)$, which occur at tree level due to the exchange of various scalar and vector leptoquarks as mentioned in the above table. Here we have summarized the observable lepton flavor non-universality parameter and the decay width of the process in the presence of leptoquark coupling.
\paragraph{}
New physics scenarios that aim to explain anomalies in the charged-current observables of semileptonic B meson decays also produce effects in the neutral-current observables of bottomonium mesons $R_ {\Upsilon(ns)}$, with $n = 1, 2, 3$. We constrain the parameter for $\Upsilon \to \tau^+ \tau^- $ using the data set given for $ b\to c \tau \bar{\nu}_\tau$ process. As the main result of our analysis which includes the various  scalar and vector leptoquarks, coupling to the standard model fermions, it is found that the lepton flavor non-universality parameter is in good agreement with the experimental result within the allowed range.  However, we found that  the influence of new physics on $R_\Upsilon$ is negligibly small.

\section{Acknowledgement}
SP would like to acknowledge the university non-net fellowship grant, DP acknowledge the support from University of Hyderabad through IoE project grant No. RC1-20-012 and AB would like to acknowledge DST INSPIRE program for financial support.



\end{document}